# Retrieval of material parameters for uniaxial metamaterials

Georgia T. Papadakis[1,†], Pochi Yeh[2] and Harry A. Atwater[1]

[1] *Thomas J. Watson Laboratories of Applied Physics, California Institute of Technology, California 91125, USA*
[2] *Department of Electrical and Computer Engineering, University of California, Santa Barbara, California 93106, USA*

[†]Corresponding author e-mail: gpapadak@caltech.edu

**Abstract**

We present a general method for retrieving the effective tensorial permittivity of any uniaxially anisotropic metamaterial. By relaxing the usually imposed condition of non-magnetic metal/dielectric metamaterials, we also retrieve the permeability tensor and show that hyperbolic metamaterials exhibit a strong diamagnetic response in the visible regime. We obtain global material parameters, directly measurable with spectroscopic ellipsometry and distinguishable from mere wave parameters, by using the generalized dispersion equation for uniaxial crystals along with existing homogenization methods. Our method is analytically and experimentally verified for $Ag/SiO_2$ planar metamaterials with varying number of layers and compared to the effective medium theory. We also propose an experimental method for retrieving material parameters using methods other than ellipsometry.

## I. Introduction

Today's nanofabrication techniques enable us to built artificial composite media, also called metamaterials (MM), with sub-wavelength unit cells, usually termed meta-atoms. MMs can be engineered to have extraordinary optical properties that cannot be found in nature, some of which are negative refractive index[1,2], reversed Doppler effect[2], epsilon near zero[3,4], super-resolution[5] (see Ref. [6] and references therein). The meta-atom arrangement, which is periodic in most cases, is a crucial parameter for controlling light propagation in the MM and thus its optical response. Specifically, uniaxial MMs that have permittivities with opposite sign along different coordinate directions, also called hyperbolic metamaterials[7] (HMMs), are attracting increasing attention because they can support negative refraction[7], hyper-lensing[8] and a high density of optical states [9,10] among other phenomena.

Provided that the wavelength of light is much larger than the unit cell of the MM, the collective response of a MM can be approximated by that of a homogeneous medium. The homogenization approximation for MMs with nanoscale dimensions in the optical regime has already been established to be accurate theoretically and experimentally and is taken here as valid[11,12,13,14,15]. An effective permittivity and permeability can then be introduced[11,16-18]. However most retrieval methods reported so far consider only normal incidence[16,17,18]. In the work by C. Menzel et al.[11], a new retrieval procedure for angle dependent effective parameters was introduced. As these authors noted in their work, the retrieved parameters are then "mere *wave* parameters rather than global *material* parameters". Thus, no direct information about the anisotropy of the MM (i.e. its effective birefringence or dichroism), a key feature for HMMs, is directly obtained. Additionally, wave parameters cannot be directly measured experimentally nor do they represent constitutive parameters of the material. To be useful effective material parameters, the retrieved parameters ought to be independent of the polarization and the angle of incidence, independent of the wave-vector (local) and of the total MM thickness in the long wavelength limit.

In this paper, we utilize the homogenization of Ref. [11] for the case of uniaxially anisotropic MMs to retrieve global effective permittivity and permeability tensors. Unlike the approaches taken by previous researchers[7-10, 12-15,19-22] we do not assume unity magnetic permeability along all symmetry directions. Instead, we use the general dispersion relations for magnetic uniaxial slabs to retrieve an effective permeability tensor, in addition to the effective permittivity tensor. Until now, metal/dielectric HMMs have been assumed to be non-magnetic[7-10, 12-15,19-22] at optical frequencies. We show by contrast that they can exhibit strong diamagnetic behavior. We show that the retrieval of the complex constitutive parameters is analytical and not subject to any numerical fitting process[12,23], and that those parameters are directly measurable with spectroscopic ellipsometry. We also propose another experimental approach to global material parameter retrieval that utilizes reflection/transmission amplitude measurements together with interferometry for phase measurements. We systematically compare our solution to the inverse problem with that of the forward problem and obtain excellent agreement for all angles of incidence.

The effective response of a MM in terms of global material parameters is often approximated with an effective medium theory. The two most widely used effective medium theories in the MMs field are the generalized effective medium approximation[24] and the Bruggeman[25] approximation. Both of them are based on field averaging over the unit-cell scale. The generalized effective medium approximation is the most commonly used method for approximating an effective permittivity tensor for HMMs[7,15,19-22,26,27]. However it is only valid for low filling fractions and does not take into account the finite thickness of the MM slab. We compare our analysis to the effective medium approximation.

This paper is structured as follows. In part II we provide a short description of the work done by Ref. [11], we introduce the additional correction for uniaxial anisotropy and present our methodology. In part III, we demonstrate the application of our method to a finite slab of Ag, as a special case of a uniaxial material with equal elements of the diagonal permittivity and permeability tensors and to Ag/SiO$_2$ multilayer MMs of different number of layers. We compare our calculations to the effective medium approximation and present experimental results obtained with ellipsometry. We also discuss the diamagnetic response of planar HMMs. In part IV we calculate the isofrequency contours for a magnetic uniaxial MM slab to highlight the importance of an accurate retrieval method for HMMs. In part V we conclude by proposing another experimental setup for direct experimental retrieval of the global material parameters of HMMs.

II. Method

**a. Homogenization**

We follow the general approach of C. Menzel et al.[11], so that once the complex reflection and transmission coefficients of a slab with finite thickness and unknown optical parameters are calculated for any angle of incidence and for transverse electric (TE) or transverse magnetic (TM) polarization, a scalar complex permittivity and permeability can be analytically calculated, under the assumption that the slab is homogeneous. However in Ref. [11], no assumption is made regarding the anisotropy of the slab. Thus, as indicated in their work, the retrieved parameters $\varepsilon_{TE}$, $\mu_{TE}$, $\varepsilon_{TM}$ and $\mu_{TM}$ are *wave* parameters and not fundamental *material* parameters. Relative to the notation adopted in Ref. [11], we interchange $\varepsilon_{TM}$ and $\mu_{TM}$, due to the correspondence of the electric and magnetic field between TE and TM polarization in homogeneous media. The effective refractive indices for the TE and TM waves respectively are defined as: $n_{TE} = \sqrt{\varepsilon_{TE}\mu_{TE}}$ and $n_{TM} = \sqrt{\varepsilon_{TM}\mu_{TM}}$.

**b. Dispersion equations for a magnetic uniaxial crystal**

Now we add the constraint that the slab of unknown parameters is uniaxial and its optical axis coincides with the direction of normal incidence ($z$ axis in the Fig. 1).

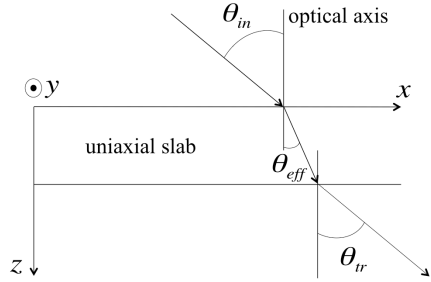

**Fig. 1**: Cross section through the MM slab, convention for the optical axis and angle of incidence

Assuming a monochromatic plane wave, the electric and magnetic field vectors are respectively $\mathbf{E}(\mathbf{r},t) = \mathbf{E}_o e^{i(\mathbf{kr}-\omega t)}$ and $\mathbf{H}(\mathbf{r},t) = \mathbf{H}_o e^{i(\mathbf{kr}-\omega t)}$. Inserting the fields into Maxwell's equations[28] gives

$$\mathbf{k} \times \mathbf{E} = \omega \vec{\mu} \mathbf{H} \tag{1}$$

$$\mathbf{k} \times \mathbf{H} = -\omega \vec{\varepsilon} \mathbf{E} \tag{2}$$

By eliminating $\mathbf{H}$ from Eq. (1)-(2), we obtain:

$$\mathbf{k} \times \vec{\mu}^{-1}(\mathbf{k} \times \mathbf{E}) + \omega^2 \vec{\varepsilon} \mathbf{E} = 0 \tag{3}$$

For a uniaxial slab, since the optical axis is aligned to the $z$ axis, the permittivity and permeability tensors are diagonal with:

$$\vec{\varepsilon} = diag(\varepsilon_{xx}, \varepsilon_{yy}, \varepsilon_{zz}) = diag(\varepsilon_o, \varepsilon_o, \varepsilon_e) \text{ and } \vec{\mu} = diag(\mu_{xx}, \mu_{yy}, \mu_{zz}) = diag(\mu_o, \mu_o, \mu_e) \tag{4}$$

where we replaced $\varepsilon_{xx} = \varepsilon_{yy}$ with $\varepsilon_o$, the ordinary permittivity of a uniaxial slab and $\varepsilon_{zz}$ with $\varepsilon_e$, the extraordinary permittivity of the slab, similarly for the permeability tensor elements $\mu_o$ and $\mu_e$. (Here we refer to $\varepsilon_o$ and $\mu_o$ as the ordinary parameters in order to be consistent with the literature of non-magnetic uniaxial materials[28], and likewise for the extraordinary parameters). Carrying out the algebra of Eq. (3), the dispersion equations for magnetic uniaxial crystals are:

Magnetically extraordinary wave: $\dfrac{k_x^2 + k_y^2}{\varepsilon_o \mu_e} + \dfrac{k_z^2}{\varepsilon_o \mu_o} = k_o^2$ (5a)

$$\text{Electrically extraordinary wave: } \frac{k_x^2 + k_y^2}{\varepsilon_e \mu_o} + \frac{k_z^2}{\varepsilon_o \mu_o} = k_o^2 \tag{5b}$$

where $k_0 = \omega/c$. Thus for a magnetic uniaxial slab the dispersion surface consists of two ellipsoids of revolution. The parameters $\varepsilon_o, \mu_o, \varepsilon_e, \mu_e$ are fundamental, angle and thickness independent, *global material* parameters. Let us now assume that the electromagnetic wave is propagating in the *xz* plane. Then, $k_y = 0$ and the magnetically extraordinary wave corresponds to electric field parallel to the *y* axis (TE polarization) while the electrically extraordinary wave corresponds to magnetic field parallel to the *y* axis (TM polarization). We emphasize here that, a HMM can be designed not only by requiring permittivities of opposite signs along different axes for the TM wave, but also by requiring permeabilities of opposite signs along different axes for the TE wave, as recently demonstrated in the microwave regime[29].

Here $k_x$ is the in-plane wave vector that is conserved above, inside and below the slab, and $k_z$ is the normal component of the wave vector in the slab. For MM slabs $k_z$ stands for the effective wave vector usually retrieved in terms of the complex reflection and transmission coefficients, the polarization and the incident angle $\theta_{in}$[11]. For TE polarization, $k_x = k_0 sin\theta_{in} = n_{TE} k_0 sin\theta_{eff}$ and $k_z = n_{TE} k_0 cos\theta_{eff}$, similarly for TM polarization. Thus, Eq. (5a) and (5b) can be associated with the wave parameters for TE and TM polarization, $\varepsilon_{TE}, \mu_{TE}, \varepsilon_{TM}$ and $\mu_{TM}$:

$$\text{Magnetically extraordinary wave: } \frac{1}{\varepsilon_{TE}(\theta_{in})\mu_{TE}(\theta_{in})} = \frac{sin^2\theta_{eff}}{\varepsilon_o \mu_e} + \frac{cos^2\theta_{eff}}{\varepsilon_o \mu_o} \tag{6a}$$

$$\text{Electrically extraordinary wave: } \frac{1}{\varepsilon_{TM}(\theta_{in})\mu_{TM}(\theta_{in})} = \frac{sin^2\theta_{eff}}{\varepsilon_e \mu_o} + \frac{cos^2\theta_{eff}}{\varepsilon_o \mu_o} \tag{6b}$$

Where $\theta_{eff}$ is the refraction angle into the slab (See Fig. 1). Since the denominators of the right hand side of Eqs. (6a) and (6b) must be angle independent, the wave parameters are angle-dependent. In the following section we verify this argument.

**c. Retrieval of ordinary parameters $\varepsilon_o, \mu_o$**

For normal incidence, $k_z = n_{TE}(\theta_{in} = 0)k_0 = n_{TM}(\theta_{in} = 0)k_0$ and from the equations above we obtain: $n_{TE}^2(\theta_{in} = 0) = n_{TM}^2(\theta_{in} = 0) = \varepsilon_0 \mu_0$. Thus $\varepsilon_o = \varepsilon_{TE}(\theta_{in} = 0) = \varepsilon_{TM}(\theta_{in} = 0)$ and $\mu_o = \mu_{TE}(\theta_{in} = 0) = \mu_{TM}(\theta_{in} = 0)$. This is expected since at normal incidence the two polarizations are degenerate. Thus, the application of the wave parameter retrieval method [11] for normal incidence yields the ordinary parameters of any uniaxial slab, illustrated in the geometry of Fig.1.

**d. Retrieval of extraordinary parameters $\varepsilon_e, \mu_e$**

To retrieve the effective extraordinary permeability we solve Eq. (5a) for $\mu_e$ where $k_z$ represents the effective wave vector for TE polarization for any oblique angle of incidence except $0^o$. To retrieve the effective extraordinary permittivity we solve Eq. (5b) for $\varepsilon_e$ where $k_z$ stands for the effective wave vector for TM polarization for any oblique angle of incidence except $0^o$. For any oblique angle of incidence , the application of Eqs. (6a) and (6b) yields angle-independent parameters, as we prove below.

### III. Application to a slab of Ag and to planar Ag/SiO$_2$ HMMs

We give here results for material parameter retrieval applied to a) a 20nm thick Ag slab, which is viewed as a special case of a uniaxial material with $\varepsilon_o = \varepsilon_e$ and $\mu_o = \mu_e$, and to planar MMs consisting of b) three, c) five and d) seven alternating layers of Ag and SiO$_2$, with thicknesses 20nm each. Planar metal/dielectric MMs were fabricated by thin film evaporation, and experimentally measured with spectroscopic ellipsometry. A five poles Drude-Lorentz (DL) model was used for the dielectric function of Ag $\varepsilon_{Ag\_DL}$[30]. The Sellmeier equation was used for the refractive index of SiO$_2$[31]. Using the transfer matrix method[28] for layered media, we calculate the complex reflection and transmission coefficients for angles of incidence $0^o - 90^o$ for TE and TM polarization.

**a. Retrieval of wave parameters $\varepsilon_{TE}$, $\mu_{TE}$, $\varepsilon_{TM}$ and $\mu_{TM}$ and ordinary parameters $\varepsilon_o$ and $\mu_o$**

We are then able to calculate $\varepsilon_{TE}$, $\mu_{TE}$, $\varepsilon_{TM}$ and $\mu_{TM}$ for different angles of incidence, shown in Fig. 2 and Fig. 3 respectively.

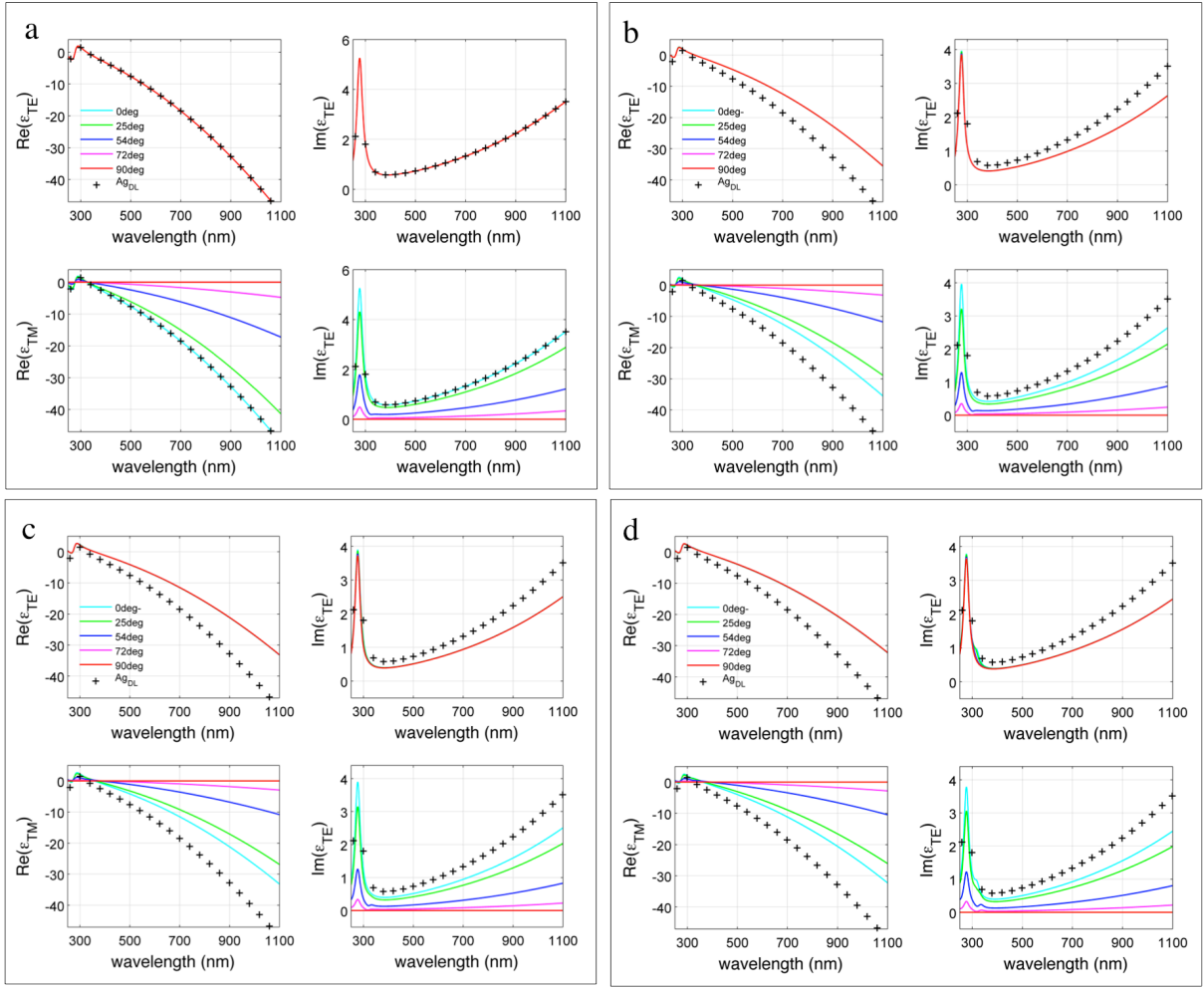

**Fig. 2:** $\varepsilon_{TE}$, $\varepsilon_{TM}$ for a) a single layer of Ag and for a MM consisting of b) 3, c) 5, d) 7 layers for different incident angles

At normal incidence (light blue curve) $\varepsilon_{TE}$ and $\varepsilon_{TM}$ coincide, as expected since the two polarizations are degenerate at $0^o$ and they represent $\varepsilon_o$ as justified in Sec. II, c. $\varepsilon_{TM}$ is clearly angle dependent. At $90^o$, $\varepsilon_{TM}$ is approaching zero. This is accompanied with a very large value of $\mu_{TM}(\theta_{in}=90^o)$ and is justified below. Contrary to $\varepsilon_{TM}$, $\varepsilon_{TE}$ exhibits almost no angle variation since the plots representing different angles converge to the same curve for all the considered number of layers. From Eq. (6a), this indicates the isotropic magnetic character of the MMs: as long as $\mu_{TE}$ remains angle independent as well, this reveals that $\mu_o \approx \mu_e$ which is shown below. For the single slab of Ag (Fig. 2a) $\varepsilon_{TE}$ for all different angles of incidence trivially converges to the same curve, the one of the 5-poles DL model of Ag: $\varepsilon_{Ag\_DL}$. For the three, five and seven layer multilayer MMs, the redshift of $\varepsilon_o$ with respect to $\varepsilon_{Ag\_DL}$ originates from the insertion of the SiO$_2$ layers between adjacent Ag layers that effectively yields a less polarizable (meta) material.

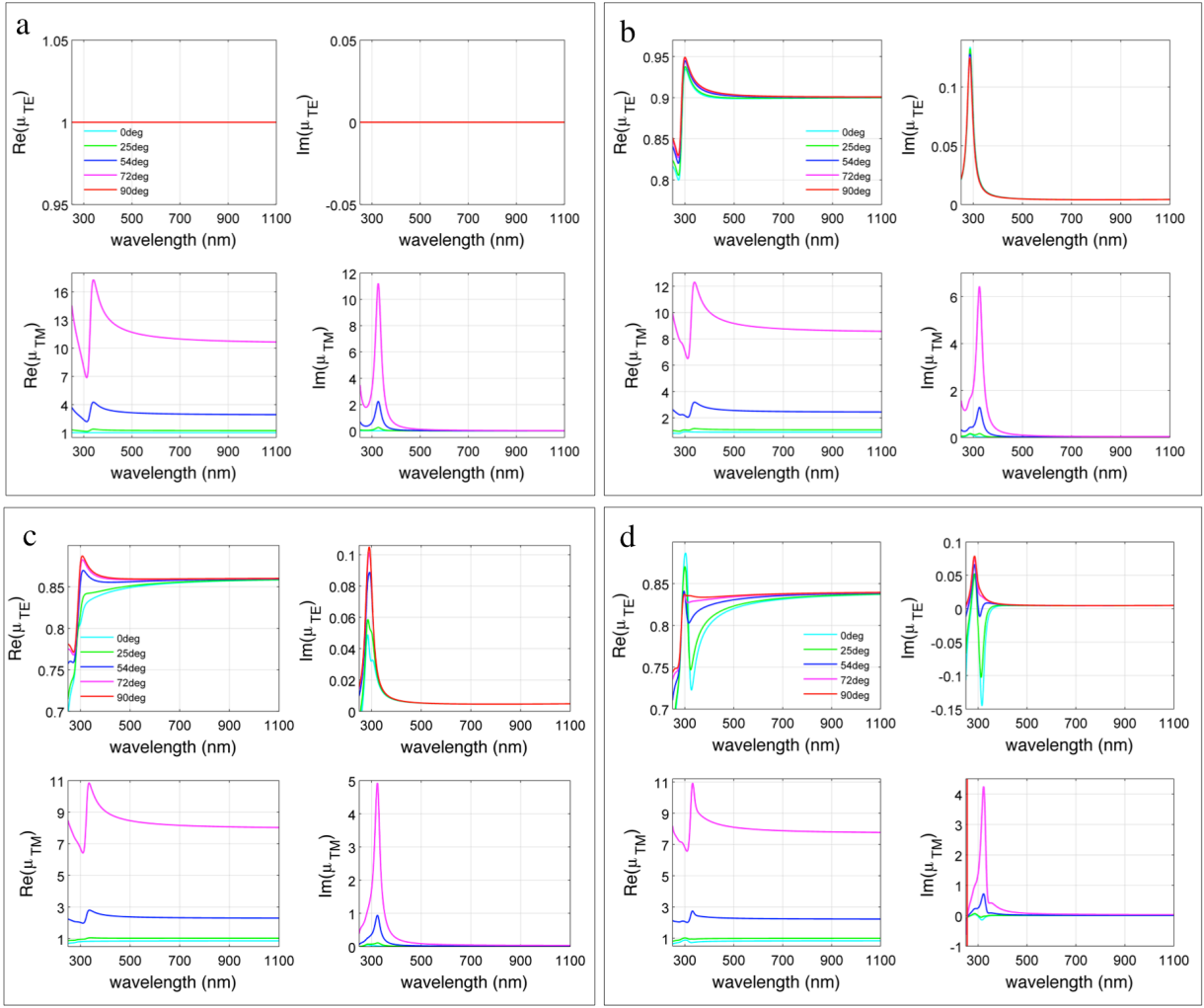

**Fig. 3:** $\mu_{TE}$, $\mu_{TM}$ for a) a single layer of Ag and for a MM consisting of b) 3, c) 5, d) 7 layers for different incident angles

The ordinary permeability $\mu_o$ is represented by $\mu_{TE}$ and $\mu_{TM}$ at normal incidence (light blue curve) as justified in Sec. II.c. For the single Ag slab in Fig. 3a, $\mu_0 = \mu_{TE/TM}(\theta_{in} = 0)$ converges to the value $1+0i$ as expected for a non-magnetic material. For the MMs investigated here, $\mu_{TE}$ has a very small angular variation compared to the variation of $\mu_{TM}$ (See axes in all plots of Fig.3 top and bottom), which implies magnetically isotropic MMs as will be proved upon calculation of $\mu_e$. The parameter $\mu_{TM}$ is strongly angle dependent for MMs with 3,5, and 7 layers, as a mere wave parameter. At $90^o$, the values of both real and imaginary part of $\mu_{TM}$ are of the order of $10^5$, and they are excluded from the plots. This agrees with the very small values of $\varepsilon_{TM}(\theta_{in} = 90^o)$ in Fig. 2. The product $\varepsilon_{TM}\mu_{TM}$ is finite and represents the square of the modal effective index for the TM wave. In Fig. 3d for the 7 layers MM, for wavelengths smaller than 350nm, we find that the imaginary parts of $\mu_0 = \mu_{TE/TM}(\theta_{in} = 0)$ and also of $\mu_{TE}$ and $\mu_{TM}$ at other angles are negative, but this is not of any concern since causality is not violated as justified in Sec. III.c.

**b. Retrieval of extraordinary parameters $\varepsilon_e$ and $\mu_e$**

Following Sec. II.d, the global material parameters $\varepsilon_e$ and $\mu_e$ for the single slab and the multilayer MMs are calculated and shown below.

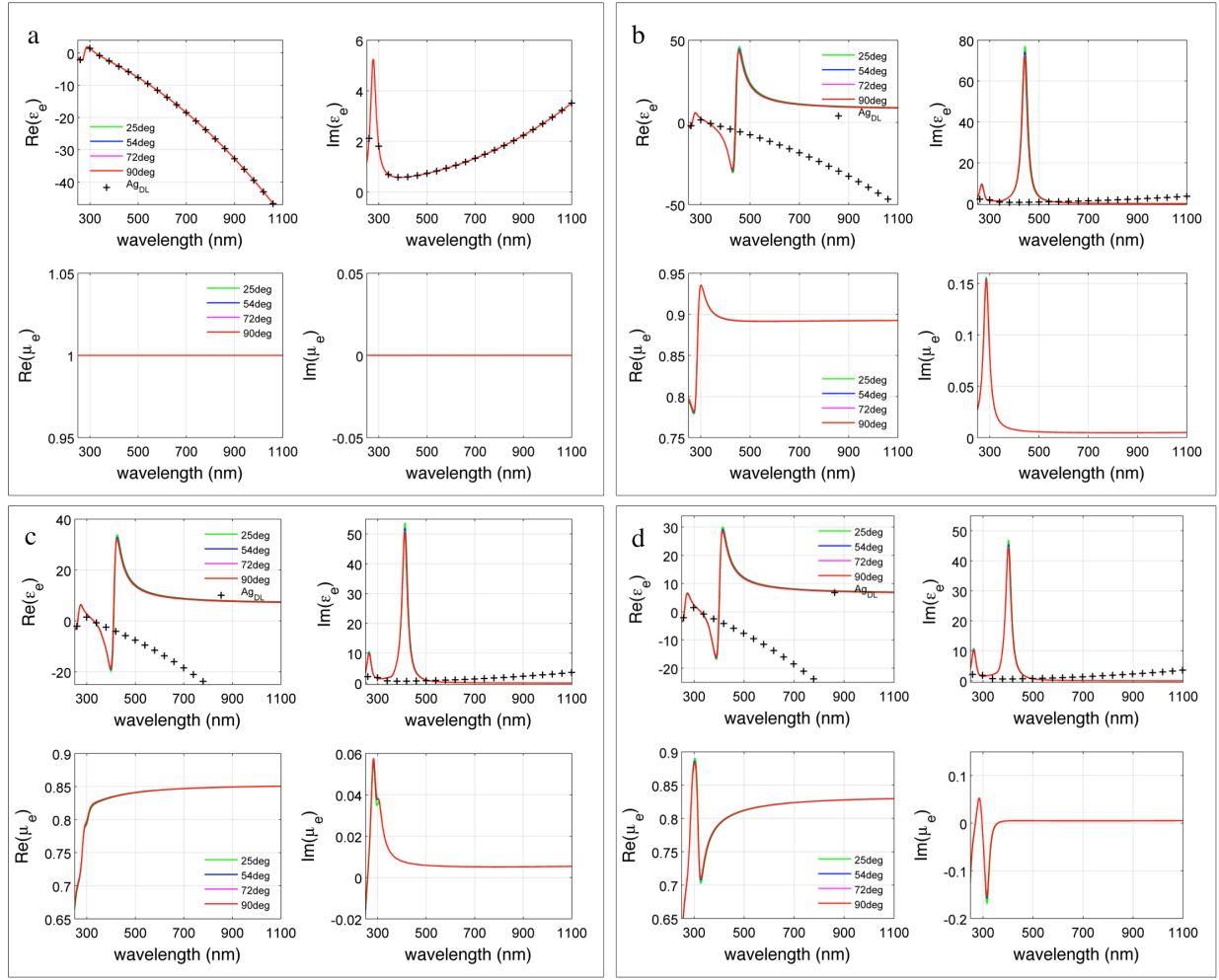

**Fig. 4:** $\varepsilon_e$, $\mu_e$ for a) a single layer of Ag and for a MM consisting of b) 3, c) 5, d) 7 layers for different incident angles

It is remarkable that for the three, five and seven layer metamaterials considered here, both $\varepsilon_e$ and $\mu_e$ exhibit negligible angular variation. They always converge to the same curve as shown above. Thus, their "intrinsic" nature as global *material* parameters, contrary to $\varepsilon_{TE}, \mu_{TE}, \varepsilon_{TM}$ and $\mu_{TM}$ -merely wave parameters-, is now obvious. For the case of the single Ag slab (Fig. 4a) $\varepsilon_e$ exactly converges to $\varepsilon_{Ag\_DL}$ and $\mu_e = 1 + 0i$ for all angles of incidence.

Concerning Fig. 2b, c, d and 4b, c and d, the mostly metallic response of these planar MMs in the $xz$ plane where $Re(\varepsilon_o) < 0$ is expected since the charge carriers are free to move in this plane. The Lorentzian-shaped $Im(\varepsilon_e) < 0$ and its Kramers-Kronig counterpart $Re(\varepsilon_e)$ of the MMs are also common features in uniaxial crystals existing in nature like calcite at mid-infrared wavelengths[32]. To conclude, for a single slab of Ag the application of the retrieval trivially yields $\varepsilon_o = \varepsilon_{TE/TM}(\theta_{in} = 0^o) = \varepsilon_e = \varepsilon_{Ag\_DL}$ and $\mu_o = \mu_{TE/TM}(\theta_{in} = 0^o) = \mu_e = 1 + 0i$ for all angles of incidence $0^o - 90^o$, yielding a consistency check for the model for the case of an isotropic non-magnetic slab.

### c. The effective permeability of Ag/SiO$_2$ HMMs and conservation of energy

The difference between $\mu_o$ and $\mu_e$ is of the order of $10^{-3}$ for all the MMs considered here, so these MMs can be considered as magnetically isotropic. As seen from Fig. 3 and 4, the real part of $\mu_o$ and $\mu_e$ can be as low as 0.825. The most diamagnetic material existing in nature is bismuth with a permeability of 0.999834. Thus, HMMs are seen to exhibit a strong diamagnetic response which can be understood as follows: the tangential component of the magnetic field induces a surface current at the Ag/SiO$_2$ interfaces which, according to Lenz's law, must create a magnetic response opposing the applied magnetic field. Regarding the negative imaginary part of $\mu_o$ and $\mu_e$ for the MM consisting of 7 alternating layers of Ag and SiO$_2$, we assure that conservation of energy is not violated because[1]:

1. Through the homogenization procedure[11], a positive imaginary part of $k_z$ was imposed. The parameters $\mu_{TE}$, $\mu_{TM}$, $\mu_o$ and $\mu_e$ are calculated through $k_z$ (through Eq. (5a) and (5b)), thus assuming a passive medium.

2. The modal effective index of the TE and TM mode is $n_{TE} = \sqrt{\varepsilon_{TE}\mu_{TE}}$ and $n_{TM} = \sqrt{\varepsilon_{TM}\mu_{TM}}$ respectively. The imaginary parts of both $n_{TE}$ and $n_{TM}$ are positive and thus assuring a passive medium.

3. The transfer matrix[28] method has been applied not only to the multilayer MMs (forward problem) but also to the equivalent effective single homogeneous slabs with the retrieved refractive indices $n_{TE}$ and $n_{TM}$ for TE and TM polarization respectively (inverse problem). The sum of the calculated transmittance and reflectance is in both cases is smaller than unity for all incident angles and all wavelengths, assuring a positive absorption coefficient.

### d. Experimental verification with spectroscopic ellipsometry

To experimentally verify our method for multilayer MMs, we fabricate MMs consisting of 3, 5 and 7 alternating layers of Ag and SiO$_2$ using evaporated thin films (thermal deposition) on fused silica. A scanning electron microscope (SEM) image of the 5 layer MM is shown in the inset of Fig. 5a. Upon performing ellipsometric measurements (WVASE® by J. A. Woollam Co., Inc.) for angles of incidence from $55^o$ up to $55^o$ with a step of $5^o$ we fit the experimental data using as a model the retrieved parameters $\varepsilon_o$ and $\varepsilon_e$. The results are shown below.

---

[1] An extended discussion in the literature regarding the issue of the negative imaginary part of the permeability of a diamagnetic MM is in agreement with our results (see Ref. 34 and references therein).

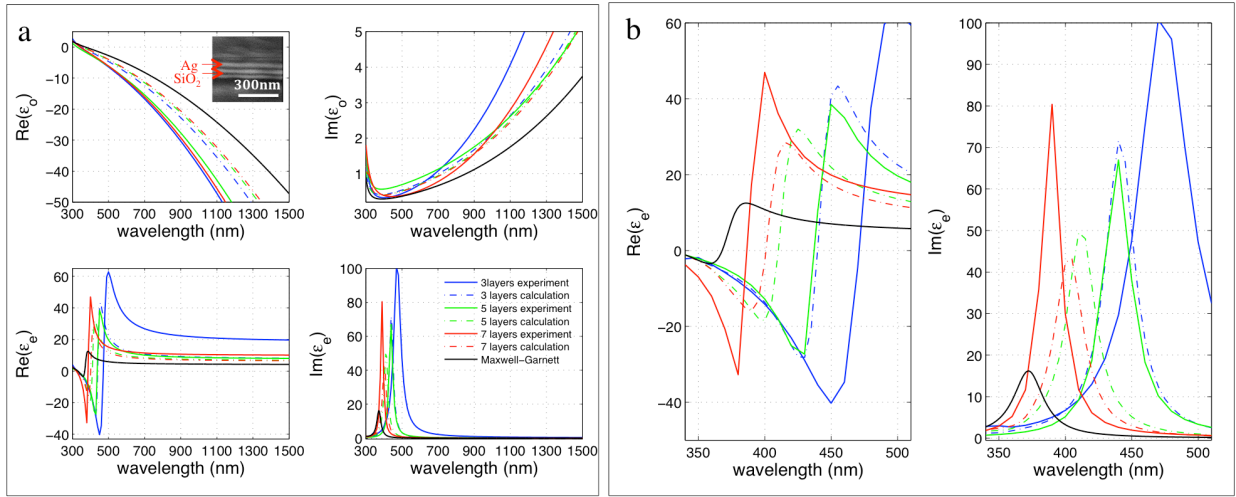

**Fig. 5**: a) Experimental (solid lines) and retrieval (dashed lines) results for $\varepsilon_o$ and $\varepsilon_e$ for planar Ag/SiO$_2$ MMs consisting of 3, 5 and 7 layers and comparison to the generalized effective medium approximation. b) same results for $\varepsilon_e$ near resonance

In the effective medium limit, the tensorial permittivity for planar periodic HMMs[7,15,19-22,24,26,27] consists of ordinary and extraordinary permittivities which are respectively $\varepsilon_o = \rho\varepsilon_m + (1-\rho)\varepsilon_d$ and $\varepsilon_e^{-1} = \rho\varepsilon_m^{-1} + (1-\rho)\varepsilon_d^{-1}$, where $\varepsilon_m$ is the metal permittivity and $\varepsilon_d$ is the dielectric permittivity and $\rho$ is the filling fraction of the metal in the unit cell. The deviation of both the experimental results and the calculations from generalized effective medium approximation is obvious in Fig. 5a. Our calculations lie inbetween the experimental curves and the effective medium theory curves. The disagreement between calculations and experiment originates from the use of bulk parameters for the index of Ag and SiO$_2$ for our calculations. Since all fabricated layers thicknesses were roughly 20nm, the surface effects on the refractive indices of the layers become important and give rise to this deviation. As the number of layers increases, the homogenization is more valid and the effective medium approximation becomes more realistic, thus, the discrepancy between both the calculations and experimental results with the effective medium decreases.

The wavelength at which the real part of $\varepsilon_e$ crosses zero, which corresponds to the center of the Lorentzian-shaped $Im(\varepsilon_e)$, blueshifts (See Fig.5b) as the number of layers increases. Specifically, for the experimentally measured $\varepsilon_e$ it is positioned at 471nm, 437nm, 386nm for 3, 5 and 7 layers respectively whereas it is positioned at 368nm for the generalized effective medium approximation. This wavelength is critical for the distinction between the spectral regions at which the MM exhibits a hyperbolic or an elliptical dispersion, giving rise to very different optical response as discussed in the next section. There is also a clear discrepancy between the amplitude $Re(\varepsilon_e)$ in the effective medium limit and our calculations and experimental results. Importantly, this affects the shape of isofrequency contours of the MM as shown in Section IV.

**e. Effect of the number of layers- Comparison to Effective Medium Theory**

We present below the summarized results of our calculations for $\varepsilon_o$ and $\varepsilon_e$ for increasing number of layers ranging from 3 to 27 layers of the Ag/SiO$_2$ multilayer MMs for comparison to effective medium theory.

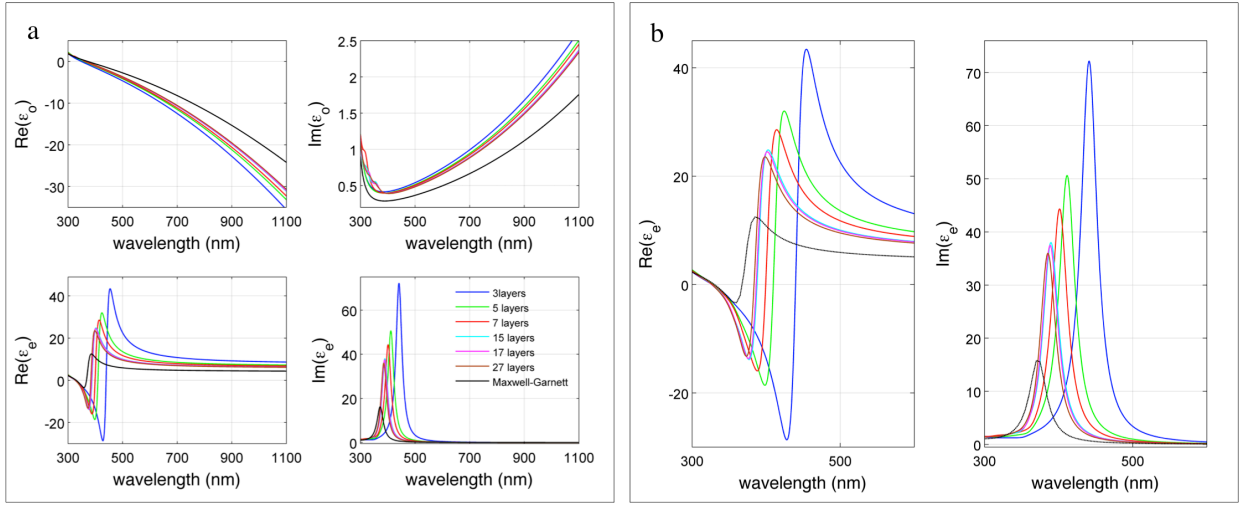

**Fig. 6**: a) $\varepsilon_o$ and $\varepsilon_e$ for planar Ag/SiO$_2$ MMs consisting of 3, 5, 7, 15, 17, 27 layers and comparison to the effective medium approximation, b) same results for $\varepsilon_e$ near resonance

As the number of layers increases, the predicted permittivities approach the effective medium approximation results, which is expected since an asymptotically large number of layers corresponds to an infinite medium[24]. However, even at 27 layers neither $\varepsilon_o$ nor $\varepsilon_e$ converge to the effective medium theory results. This suggests that the results of our retrieval method give a better prediction for the effective response of any planar HMM with a finite number of layers and provides a more appropriate model for spectroscopic ellipsometry than effective medium model.

## IV. Comparison of Isofrequency Surfaces from Retrieved Material Parameters and Effective Medium Model

The effective k-space or isofrequency contours (isofrequency contour) of MMs are of crucial importance for describing its response to single-frequency excitation. Using Eq. (5a) and (5b) we show below the isofrequency contours of the 7 layer Ag/SiO$_2$ HMMs, given the values of $\varepsilon_o$, $\varepsilon_e$, $\mu_o$ and $\mu_e$ obtained in the preceding section. We choose the 7 layered MM because the assumption of a homogeneous slab becomes more appropriate as the number of layers increases and because it is in the closest agreement to the effective medium, say, as compared to the 3 layer and 5 layer MMs. In the effective medium approximation, the permeability in different axes is usually assumed to be unity, and in this limit Eq. (5a) and (5b) become: $(k_x^2 + k_z^2) = \varepsilon_o k_o^2$ and $(k_x^2/\varepsilon_e + k_z^2/\varepsilon_o) = k_o^2$ respectively, which are commonly used to describe HMMs[7,10,15,20-22,26]. We are only taking account of the real parts of the denominators of $k_x$ and $k_z$ as their imaginary parts can be directly translated to a complex frequency instead of a complex k-space [33].

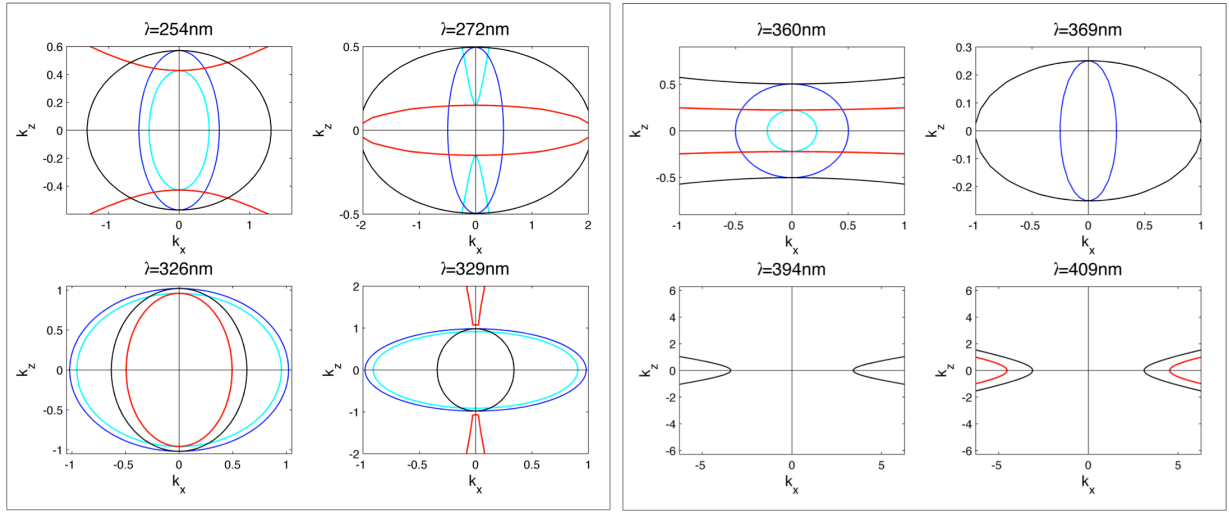

**Fig. 7**: Effective k-space for a HMM of 7 layers of Ag/SiO$_2$ and comparison to generalized effective medium theory. Results of the retrieval are shown with light blue color for TE polarization and with red for TM polarization. The generalized effective medium approximation is shown with dark blue for TE polarization and black for TM polarization

The discrepancy between the effective medium theory and our analytical calculation gives rise to isofrequency contours with different shapes, as it can be seen comparing the dark blue curves to the light blue ones (TE polarization) in Fig. 7, and the red curves to the black curves (TM polarization) respectively. Not only does the surface area enclosed by the isofrequency contour (*which is proportional to the total number of available optical states*[9]*)* differ between our calculations and the generalized effective medium approximation, at some wavelengths even the type of the dispersion changes shape from elliptical to hyperbolic for the TM wave. For example, at 254nm and at 329nm, the effective medium approximation predicts elliptical dispersion (black curves), whereas our calculations predict hyperbolic dispersion (red curves). It is worth highlighting the change of the shape of the isofrequency contour from elliptical to hyperbolic, for our calculations, between the 326nm case and the 329nm case in Fig.7 (bottom left). Within 3nm, the material transits from elliptical to hyperbolic.

For TE polarization the isofrequency contours are almost circular revealing the very weak magnetic anisotropy of the HMMs (See light blue lines). Since the imaginary parts of both products: $\varepsilon_e \mu_o$ and $\varepsilon_o \mu_o$ are omitted, when their real parts are negative Eq. (5a) is only satisfied for imaginary $k_x$ and $k_z$, which is translated to exponential decay of the TE wave inside the MM. This situation is similar to a metal that becomes a perfect reflector for frequencies below the plasma frequency. According to our calculations, this effectively metallic response for the TE wave occurs for wavelengths above 362nm which is the reason why there is no light blue curve in the figures above for the wavelengths 369nm, 394nm and 409nm. However, in the effective medium approximation it occurs at 372nm (See Fig.5a) which is the reason why for the wavelengths 394nm and 409nm in the figures above there is no dark blue curve.

A similar situation can occur for the TM wave when both $\varepsilon_o \mu_o$ and $\varepsilon_e \mu_o$ are negative (See Eq. (5b)). For this MM design, the effective medium approximation does not predict such a spectral region (See Fig.5a). However according to our calculations and to the experimental results for the fabricated MMs shown in Fig. 5, there exists a region where both $\varepsilon_o \mu_o$ and $\varepsilon_e \mu_o$ are negative: from 362nm, where the product $\varepsilon_o\mu_o$ crosses zero, up to 401nm, where the product $\varepsilon_e\mu_o$ returns to positive values. Thus, for these wavelengths, there exists an effective omnidirectional band gap for the TM wave in the MM. This is why there is no red isofrequency contour curve for the wavelengths of 369nm and 394nm in Fig.6.

This is also demonstrated in Fig. 8 where we depict the effective band structure of the MM for TM polarized light along the $k_x$ and along the $k_z$ axis. In both figures we are comparing our calculations (blue color) to the band structure from the effective medium approximation (red color). The disagreement between our calculations and the effective medium theory gives rise to different predictions for the optical band gaps of MMs, which is important for future MM applications[10,21,22].

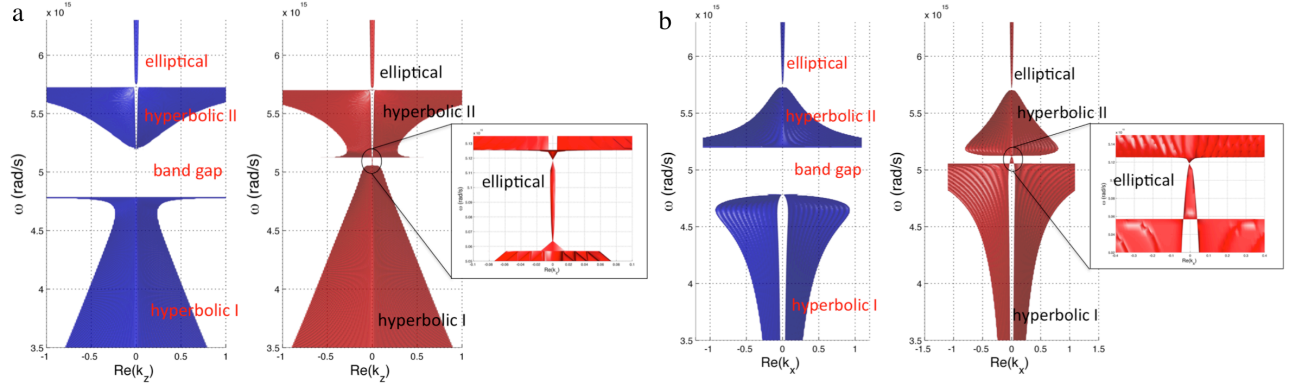

**Fig.8:** Effective band structure for 7 layer Ag/SiO$_2$ MM along a) $k_x$ axis and b) $k_z$ axis. Blue- results of the retrieval, Red- generalized effective medium theory

The low frequency region labeled as hyperbolic I region corresponds to the case where $\varepsilon_o\mu_o$ is negative and $\varepsilon_e\mu_o$ is positive. Within the spectral region of the figures (which corresponds to free space wavelengths ranging from 538nm to 300nm) the MM also supports a region where $\varepsilon_e\mu_o$ is negative and $\varepsilon_o\mu_o$ is positive ($5.2 \cdot 10^{15}$ rad/s up to $5.7 \cdot 10^{15}$ rad/s), which we note as hyperbolic region II. In this spectral region we have slow light propagation in the $z$ direction, as shown in Fig. 8a. This effect cannot be observed in natural uniaxial materials due to the positive sign of both $\varepsilon_e\mu_o$ and $\varepsilon_o\mu_o$. In the regions of hyperbolic dispersion type I or II, the bands enclose a large volume of phase space –corresponding to a very large density of optical states- compared to the elliptical dispersion seen at high frequencies, where both $\varepsilon_o\mu_o$ and $\varepsilon_e\mu_o$ are positive and approaching the characteristics of natural materials.

The origin of the large gap in the dispersion surface of Fig.8a is the negative sign of both $\varepsilon_o\mu_o$ and $\varepsilon_e\mu_o$ for these frequencies (See Fig 5a, b). The band gap is also present in Fig. 8b, indicating an omnidirectional band gap of the MM. However a similar gap does not exist in the effective medium approximation, as shown in the insets. To the contrary, there exists a small locus of $\vec{k}$ points for which there are allowed energies, due the absence of a region for which both $\varepsilon_o\mu_o$ and $\varepsilon_e\mu_o$ are simultaneously negative. In order to investigate quantitatively the nature of this band gap where propagation is forbidden in the lossless limit, one must take into account the complex nature of Eq. (5a). We do so by calculating the complex $k_z$ given the complex retrieved parameters $\varepsilon_o$, $\mu_o$ and $\varepsilon_e$. We show below a diagram for the real and imaginary parts of $k_z$ normalized to the free space wave vector $k_o$ as a function of the wavelength and of the normalized incident tangential wave vector $k_x/k_o$. As it can be seen from Fig. 9a, the hyperbolic regions are only accessible for large tangential wave-vectors. From Fig. 9b (inset), it is obvious that in the region where $\varepsilon_o\mu_o$ and $\varepsilon_e\mu_o$ are negative ranging from 362-401nm, the real part of the normalized normal wave vector is negligible compared to the imaginary part, verifying our argument that this is a real omnidirectional band gap that the generalized effective medium theory does not predict.

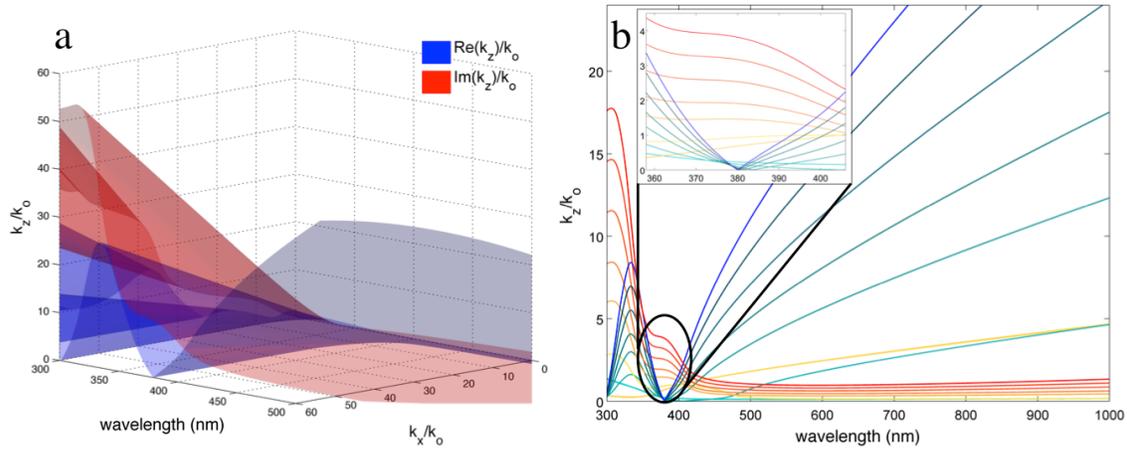

**Fig.9:** a) Real-blue and imaginary-red parts of effective normalized $k_z$ as a function of the tangential normalized wave vector $k_x$ and the wavelength. b) Projected version of Fig.9a: darker blue curves correspond to real part of normalized $k_z$ for increasing normalized $k_x$, darker red curves correspond to imaginary part of normalized $k_z$ for increasing normalized $k_x$

Finally, the very small gap at the high frequency region, around $5.7\cdot 10^{15}$ rad/s in both Fig. 8a and 8b corresponds to the transition from an elliptical dispersion to type II hyperbolic dispersion. In other words, it corresponds to the wavelength at which $\varepsilon_e\mu_o$ crosses zero. The discontinuity in the band structure comes from the fact that there is no way for a continuous isofrequency contour to transit from an elliptical shape to a hyperbolic one. This is analogous to the Lifshitz transition in electronic systems and to Van Hove singularities[22] in the density of states. Finally, from Fig. 8b, we note the negative phase velocity of the extraordinary wave in the $x$ direction in the hyperbolic II region, which is in accord with the observation of plasmonic modes with negative propagation constant in metal/dielectric waveguides for frequencies above the plasma frequency of the metal[35]. The difference here is that, instead of a natural metal we have an effective material with metallic properties above $5.2\cdot 10^{15}$ rad/s, which corresponds to the wavelength at which $\varepsilon_o\mu_o$ becomes negative.

## V. Experimental Material Parameter Retrieval

### a. Summary

Our approach to the retrieval problem for inhomogeneous uniaxial MMs is based on replacing the actual system by a homogeneous uniaxial effective medium characterized by $\varepsilon_o$, $\mu_o, \varepsilon_e, \mu_e$. The first step in our approach is to obtain the physical system's complex reflection and transmission coefficients for the two polarizations. When homogenization is valid, i.e. in the long wavelength regime, only two incident angles suffice, given that the optical axis of the MM coincides with the direction of normal incidence: for normal incidence scattering, wave parameter retrieval yields $\varepsilon_o$ and $\mu_o$. The transmission and reflection coefficient at an oblique incident angle for both TE and TM waves are sufficient for the wave parameter retrieval to yield the effective normal wave vector in the MM: $k_{zTE}(\theta_{in} \neq 0^o)$ and $k_{zTM}(\theta_{in} \neq 0^o)$ respectively. These two parameters, along with the use of Eq. (5a) and (5b), can be used to obtain $\varepsilon_e$ and $\mu_e$.

### b. Interferometry and spectrometry

For direct measurements of the material parameters, we propose an experimental protocol that involves use of an interferometric setup in reflection and in transmission for the measurement of the phase of both reflection and transmission coefficients, together with a spectrometer for the simultaneous measurement of their amplitudes, at oblique incidence. Then, these three measurements (reflection, transmission and phase) suffice for the determination of all four global material parameters $\varepsilon_o$, $\mu_o, \varepsilon_e$ and $\mu_e$: one at normal incidence, and two at an oblique indicent angle: one for TE and one for TM polarization

### c. Spectroscopic ellipsometry

Our method theoretically predicts the ordinary and extraordinary permittivity and permeability for uniaxial metamaterials and can serve as an excellent set of model parameter inputs for fits to spectroscopic ellipsometry measurements of HMMs. Specifically in Sec. III.d we showed that by utilizing our result as a uniaxial parameter model to fit the experimental data from ellipsometry, we were able to experimentally retrieve the effective tensorial permittivity of HMMs consisting of 3, 5 and 7 layers of 20nm Ag and $SiO_2$ and get better agreement than for effective medium theory.

## VI. Conclusion

We present a method for optical characterization of uniaxial metamaterials. The method can retrieve the complex elements of the permittivity and permeability tensors of any metamaterial with uniaxial anisotropy. The retrieved parameters are proved to be angle independent and thus constitute true material parameters with negligible angular dispersion in the long wavelength limit. We studied theoretically and experimentally the effect of a finite and small number of layers in planar metal/dielectric metamaterials on their effective parameters and compared results to effective medium theory. We note the importance of an accurate analytical retrieval method that improves upon existing effective medium approximations for the isofrequency surfaces of HMMs and specifically calculated an effective band gap for the electrically extraordinary wave that is not predicted by the effective medium theory. Finally, we propose two experimental protocols for retrieval of the complex material parameters for hyperbolic metamaterials. Our method is not limited to multilayer metamaterials nor to periodic structures. For example, it can be applied to a system of metallic nanorods in a dielectric host as long as the two directions perpendicular to the rods are equivalent, i.e. the structure is uniaxial. Notably, our method provides the means for the investigation of a design of a magnetic HMM and magnetic topological transitions in the optical regime, similar to the recent work in the microwave regime[34].

**Acknowledgments**

We acknowledge fruitful discussions with Prof. E. N. Economou and Dr. Tom Tiwald for help with ellipsometric modeling. This work was supported by the Multidisciplinary University Research Initiative Grant (Air Force Office of Scientific Research MURI, FA9550-12-1-0488). G. Papadakis acknowledges support by a National Science Foundation Graduate Research Fellowship.

**Author Information**

Corresponding author email: gpapadak@caltech.edu

Notes: The authors declare no competing financial interest